\documentclass[12pt]{aastex}
\usepackage{emulateapj5}

\newcommand{\HI}{\ion{H}{1}}

\newcommand{\ha}{H$\alpha$}
\newcommand{\Hp}{H$^+$}
\newcommand{\mh}{\hbox{H$_2$}}
\newcommand{\CII}{\ion{C}{2}}
\newcommand{\NI}{\ion{N}{1}}
\newcommand{\NII}{\ion{N}{2}}

\newcommand{\OVI}{\ion{O}{6}}

\newcommand{\SII}{\ion{S}{2}}

\newcommand{\PII}{\ion{P}{2}}

\newcommand{\ArI}{\ion{Ar}{1}}

\newcommand{\FeII}{\ion{Fe}{2}}
\newcommand{\FeIII}{\ion{Fe}{3}}
\newcommand{\Wlam}{\hbox{W$_{\lambda}$}}
\newcommand{\kms}{\hbox{${\rm km\:s^{-1}}$}}
\newcommand{\vlsr}{\hbox{${\rm V}_{\rm LSR}$}}
\def\cmm#1{\hbox{${\rm cm^{#1}}$}}

\def\tdex#1{\hbox{$\times10^{#1}$}}

\shortauthors{Murphy et al.}
\shorttitle{{\it FUSE} Spectroscopy of HVC Complex C}

\begin{document}

\title{{\it FUSE} Spectroscopy of High Velocity Cloud Complex C}

\author{E. M. Murphy\altaffilmark{1}, K. R. Sembach\altaffilmark{1},
B. K. Gibson\altaffilmark{2},}
\author{J. M. Shull\altaffilmark{2,3}, B. D. Savage\altaffilmark{4}, 
K. C. Roth\altaffilmark{1},}
\author{H. W. Moos\altaffilmark{1}, J. C. Green\altaffilmark{2}, 
D. G. York\altaffilmark{5}, and B. P. Wakker\altaffilmark{4}} 

\altaffiltext{1}{Department of Physics \& Astronomy, The Johns Hopkins
University, Baltimore, MD 21218}

\altaffiltext{2}{Center for Astrophysics and Space Astronomy,
Dept. of Astrophysical and Planetary Sciences,
University of Colorado, Boulder, CO 80309}

\altaffiltext{3}{Also at JILA, University of Colorado and National
Institute of Standards and Technology, Boulder, CO, 80309} 

\altaffiltext{4}{Department of Astronomy, University of Wisconsin,
Madison, WI 53706} 

\altaffiltext{5}{Astronomy \& Astrophysics Center, University of
Chicago, Chicago, IL 60637}

\begin{abstract}

We present {\it Far Ultraviolet Spectroscopic Explorer} ({\it FUSE})
observations of the sightline toward the Seyfert 1 galaxy
\objectname[]{Markarian 876}, which passes through high velocity cloud
(HVC) complex~C.  This sight line demonstrates the ability of {\it
FUSE} to measure ionic absorption lines in Galactic HVCs.  High
velocity absorption is clearly seen in both members of the \OVI\
doublet.  This is the first detection of \OVI\ in a neutral hydrogen
HVC.  One component of HVC complex~C is resolved in multiple \FeII\
lines from which we derive N(\FeII)/N(\HI)=0.48 (Fe/H)$_{\sun}$.  This
value of N(\FeII)/N(\HI) implies that the metallicity of complex~C
along this sightline may be higher than that along the Mrk 290
sightline (0.1 solar) found by \citet{wakker99}.  On the other hand,
if the metallicity of complex~C is also 0.1 solar along this line of 
sight, the observed value of N(\FeII)/(N(\HI) suggests there may be a significant amount of \Hp\ along the line of sight.
In any case, little, if any, iron can be
depleted into dust grains if the intrinsic metallicity of complex~C is
subsolar.  Absorption from complex~C is also seen in \CII, \NI, and
\NII, and upper limits based on non-detections can be determined for
\ArI, \PII, and \FeIII.  Although molecular hydrogen in the Milky Way
is obvious in the {\it FUSE} data, no \mh\ absorption is seen in the
high velocity cloud to a limit N(\mh)$<2.0\times10^{14}\;\cmm{-2}$.
Future {\it FUSE} observations of extragalactic objects behind
Galactic high velocity clouds will allow us to better constrain models
of HVC origins.
\end{abstract}


\section{Introduction}

Despite significant observational and theoretical efforts over nearly
40 years of study, the nature of Galactic \HI\ high velocity clouds
(HVCs) is still a mystery.  With the launch of the {\it Far
Ultraviolet Spectroscopic Explorer} ({\it FUSE}), a new and important
portion of the electromagnetic spectrum is now available to study
HVCs.  The far-ultraviolet provides a wealth of atomic, ionic, and
molecular spectral lines that can be used to probe the physical
conditions in HVCs \citep{sembach99}.  With its high resolution and
large effective area, {\it FUSE} has the ability to probe HVCs along
multiple lines of sight toward extragalactic objects.  This paper
presents an analysis of such a sightline and demonstrates the ability
of FUSE to significantly contribute to our understanding of HVCs.

The sight line to Mrk 876 ($l=98\fdg27$, $b=40\fdg38$) passes through
HVC complex~C \citep{wakwoer91}.  \citet{wakker99} suggest the
distance to complex~C is in the range of 5-25 kpc, with ${\rm D}\sim10$ kpc
as the most likely value. 
They derive a mass for complex~C of $6\times10^{6}\;{\rm
M}_{\sun}\;({\rm D}/10\;{\rm kpc})^{2}$ which, combined with the
observed radial velocity, implies a mass influx of $0.08-0.19\; ({\rm
D}/10\;{\rm kpc})\;{\rm M}_{\sun}\;{\rm yr}^{-1}$.  The
metallicity of complex~C is in the range 0.1 to 0.6 solar
\citep{wakker99, gibson00} and may be spatially variable toward
multiple sightlines separated by more than 10\degr.  The metallicity 
of HVCs can be used to discriminate between the current theories for
their origins \citep{vanwoerden99}.

Along the Mrk 876 sightline, a 21 cm spectrum taken with the Effelsberg 100
m telescope with a 9\farcm7 beam shows two components at
$\vlsr=-172\;\kms$ and $\vlsr=-133\;\kms$ with
N(\HI)$=(4.1\pm0.8)\times10^{18}\;\cmm{-2}$ and
N(\HI)$=(19.2\pm0.8)\times10^{18}\;\cmm{-2}$, respectively.  An
NRAO\footnote{The National Radio Astronomy Observatory is a facility
of the National Science Foundation, operated under cooperative
agreement by Associated Universities, Inc.} 140 Foot Telescope (43
meter) \notetoeditor{I have capitalized 140 Foot Telescope since this
is the proper name used by NRAO for the telescope} observation with a
21\arcmin\ beam gives a column density that is 2.2 times higher
than the Effelsberg spectrum for the $-$172 \kms\ component while both 
telescopes give identical results for the $-$133 \kms\ component.
The difference between the 9\farcm7 and 21\arcmin\ results
probably arises from small scale structure in the HVC, since it
is much larger than any expected statistical
uncertainties or calibration differences.  We have used the 
Effelsberg spectrum in the analysis that follows.

An \ha\ spectrum obtained by M. Haffner (private communication) with
the WHAM instrument \citep{reynolds98} indicates that the \ha\
intensity associated with complex~C toward Mrk 876 is $<0.02$ R (1 R
is 10$^6$ photons$\,\cmm{-2}\,(4\,\pi\,{\rm sr})^{-1}$) averaged over
the 1\degr\ beam of the instrument.  Complex~C is clearly detected in
other directions (Wakker et al.\ 1999; Tufte et al.\ 2000, in
preparation) which implies a patchy distribution of N(\Hp).  Since
both the distance to complex C and the geometry of the emitting region
are unknown, and there may exist small scale structure in the large WHAM
beam, no meaningful limit can be placed on N(\Hp) at this time.

\section{{\it FUSE} Observations}

For a description of the {\it FUSE} satellite, its operation, and
observing modes see \citet{moos00} and \citet{sahnow00}.  The Mrk 876
dataset (P1073101) consists of 10 consecutive exposures in the
$30\arcsec\times30\arcsec$ apertures, resulting in 52 ksec of on-target
integration.  At the time of the observations (16 October 1999) the
spectrograph was not yet aligned or focused.  The target appears in
both LiF channels and one SiC channel.  The data were passed through
the standard CALFUSE pipeline, which removed the spectral motion,
performed a background subtraction, removed the geometric distortions,
extracted the spectra, and performed wavelength and flux calibrations.
Event bursts were removed from the data by hand.  Coaddition of the
channels was not attempted due to the preliminary nature of the
wavelength calibration.  Typically, the equivalent width of an
absorption line was measured separately for each channel and averaged.
The typical spectral resolution was $\lambda/\Delta\lambda\approx 12000$.
The average S/N is 14:1 per resolution
element in the LiF1 channel, 17:1 in the LiF2 channel, and 8:1 in the
SiC1 channel.

\section{First Detection of an \ion{O}{6} HVC}

\begin{figure*}
\plotone{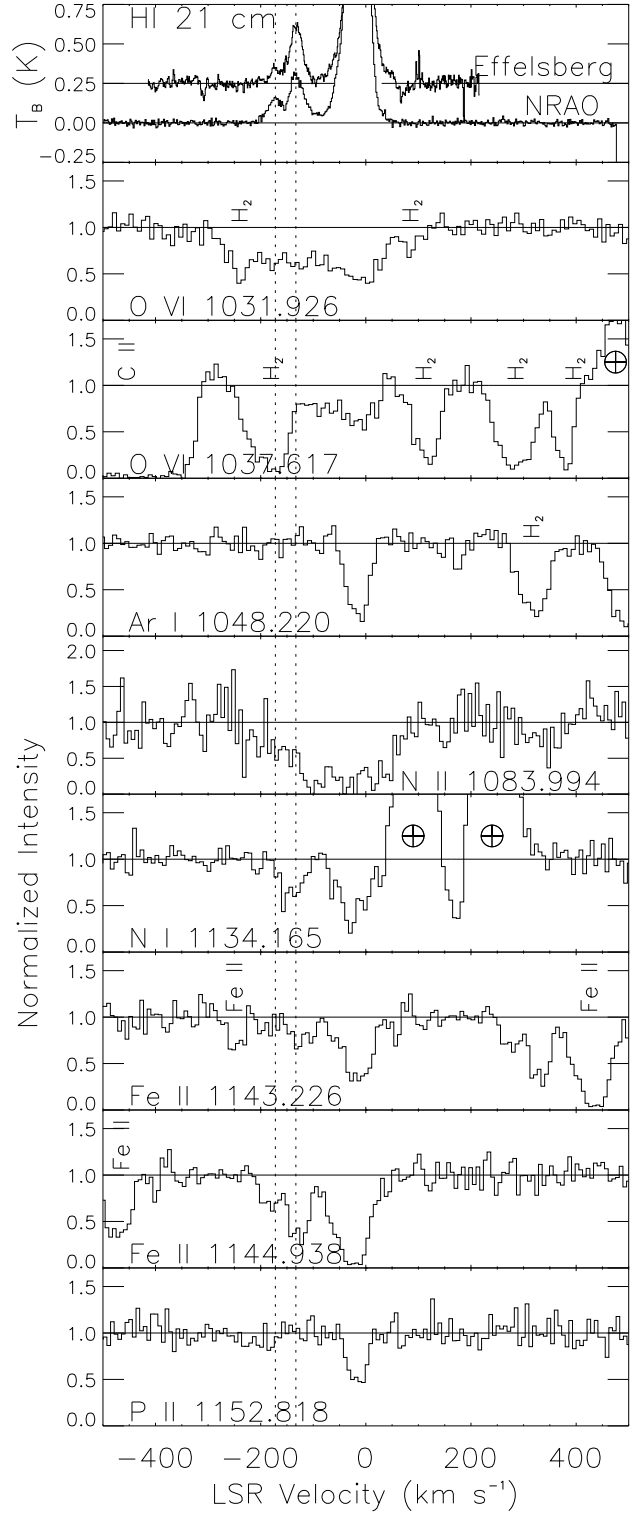}
\caption{Ionic absorption lines as measured by {\it FUSE}
along the line of sight to Mrk 876.  For species that are covered by
two or more channels, only one channel is shown.  The wavelength scale
was adjusted separately for each line by aligning the low-velocity
absorption.  Absorption from HVC complex C is clearly seen in the
\OVI\ 1031.93 \AA\ line and \FeII\ 1144.94 \AA\ line.
\label{fig1}}
\end{figure*}

The \OVI\ lines are shown in Figure 1.  Absorption from HVC complex~C
is clearly present at negative velocities as high as $-215\;\kms$.
\OVI\ is an excellent tracer of hot gas ($2-5\times10^{5}$ K) and is
not generally produced by photoionization.  Beyond $-215\;\kms$ the
\OVI\ 1031.93 \AA\ line is blended with the Galactic \mh(6-0) P(3)
line at 1031.19 \AA.  Fortunately, the \mh\ line is narrow and was
easily removed from the spectrum.  Integrating the profile between
$-215\;\kms$ and $-100\;\kms$ we find and equivalent width of
$146\pm14$ m\AA\ which implies N(\OVI)=$(1.5
\pm 0.2)\times10^{14}\;\cmm{-2}$ if the line is optically thin.  High
velocity absorption is also seen in the \OVI\ 1037.62 \AA\ line;
however, it is blended with the strong Galactic
\mh(5-0) R(1) line at velocities more negative than $-140$ \kms, making a 
column density measurement impossible.

The amount of hot gas implied by the \OVI\ detection can be
significant if the metallicity of complex~C is low.  The models of
\citet{sutherland93} give the ionization fraction of
\OVI\ as function of temperature for gas in collisional
equilibrium. At the peak of the \OVI\ ionization fraction (22\% at
280,000 K), the amount of hot gas is N(H)$>\:7\tdex{17}\;\cmm{-2}\;({\rm
O/H})_{\sun}/({\rm O/H})_{\rm HVC}$.

A discussion of the possible interpretations for the detection of
high-velocity \OVI\ in complex C and other HVCs is presented by 
\citet{sembach00}. 

\section{Metal lines in Complex C}

Metal line absorption associated with the $-133\;\kms$ \HI\ component
(and, in some cases, the $-172\;\kms$ component) is seen in several
\FeII\ lines, \CII, \NI, and \NII. Upper limits can be set for
\ArI, \PII, and \FeIII.  Example absorption lines are
presented in Figure 1. 
Table~1 lists the measured equivalent widths, column densities and
derived abundances relative to the solar values. The errors are a
combination in quadrature of the statistical error, based on the S/N
ratio of the spectrum, and systematic errors calculated as the rms
value of measurements made by different authors, by using different
detector segments, different continuum placement, and by using subsets
of the data (e.g. night-only). 

Five HVC \FeII\ lines from the $-133\,\kms$ component are detected,
from which a curve of growth can be derived.  We minimize
$\chi^2$(N,$b$)=[\Wlam(observed)$^2-$\Wlam(N,$b$)$^2$], and find
N=3.0$\pm$1.2\tdex{14}\,\cmm{-2}\ and $b$=12.1$\pm$5.6\,\kms. All
observed values of \Wlam are then within 1$\sigma$ of the expected
value. We will use this $b$-value below to convert equivalent width to
column density for other ions (excluding \OVI).  The $-$172 \kms\
component is also seen in absorption in
\FeII\ $\lambda$1144.94.  The \FeIII\ $\lambda$1121.98 line from the
$-133\,\kms$ component is absent, yielding N(\FeIII)/N(\FeII)$\,<\,$0.22. 

\NII\ $\lambda$1083.99 high-velocity absorption can clearly be seen in the
SiC1 channel; however, the individual components are blended and the
spectrum has low S/N. Integrating the line and apparent column density
profiles between $-$160 and $-$100\,\kms\ yields an equivalent width
$>$120\,m\AA\ and a column density of
N(\NII)$>1.0\tdex{14}\;\cmm{-2}$.  If we calculate the column density
assuming the $b$ value found above for iron, the result is a factor of
three higher.  The high-velocity component of \NI\ $\lambda$1134.17 is
blended with low-velocity \FeII\ $\lambda1133.67$.  Using a curve of
growth for the low velocity gas, we derive \Wlam(\FeII\
$\lambda$1133.67)=67$\pm$7\,m\AA.  The measured width of the blend is
94$\pm$7\,m\AA, so that 27$\pm$10\,m\AA\ can be attributed to
high-velocity \NI.  This matches, to within the errors, the column
density of \NI\ $\lambda$1199.55 found by \citet{gibson00} toward Mrk
876 using the Space Telescope Imaging Spectrograph on the {\it Hubble
Space Telescope}.

\CII\ $\lambda$1036.34 is clearly present in both components of the HVC
and is strongly saturated.  For \PII\ $\lambda$1152.82 and \ArI\
$\lambda$1048.22 only $3\sigma$ upper limits of 25\,m\AA\ can be set
at this time.

\section{Metal abundances in complex C}

The observed ratio N(\FeII)/N(\HI)$\sim0.5$ (Fe/H)$_{\sun}$ is
unexpected given that every previously studied sightline through cool,
warm, or halo gas has shown iron depleted by at least a factor 3
\citep{savage96}.  If the intrinsic metallicity of complex~C is
subsolar, then little, if any, iron can be depleted into dust grains.
The iron abundance is also higher than the value of S/H$\sim$0.1
(S/H)$_{\sun}$ found by \citet{wakker99} along the Mrk 290 sightline
through complex~C.  Gibson et al. (2000) have measured \SII\
absorption along the lines of sight to Mrk 817, where
N(\SII)/N(\HI)=0.3 (S/H)$_{\sun}$, and Mrk 279, where
N(\SII)/N(\HI)$\sim$0.6 (S/H)$_{\sun}$.  They find that assuming the
presence of \Hp\ along the line of sight is insufficient to reconcile
their observations with the metallicity found by \citet{wakker99}.
Instead, they believe that the metallicity of complex~C is spatially
variable with metallicities ranging from $0.1-0.6$ solar.  The low
abundance of argon (N(\ArI)/N(\HI)$<0.13$ (Ar/H)$_{\sun}$) toward Mrk
876 is probably a photoionization effect; \ArI\ has a photoionization
cross-section about 10 times larger than that of hydrogen
\citep{sofia98}.  Upcoming {\it FUSE} observations of the Mrk 290, Mrk
817, and Mrk 279 sightlines should help resolve these issues.

\section{Limits on Molecular Hydrogen}

\begin{figure*}
\plotone{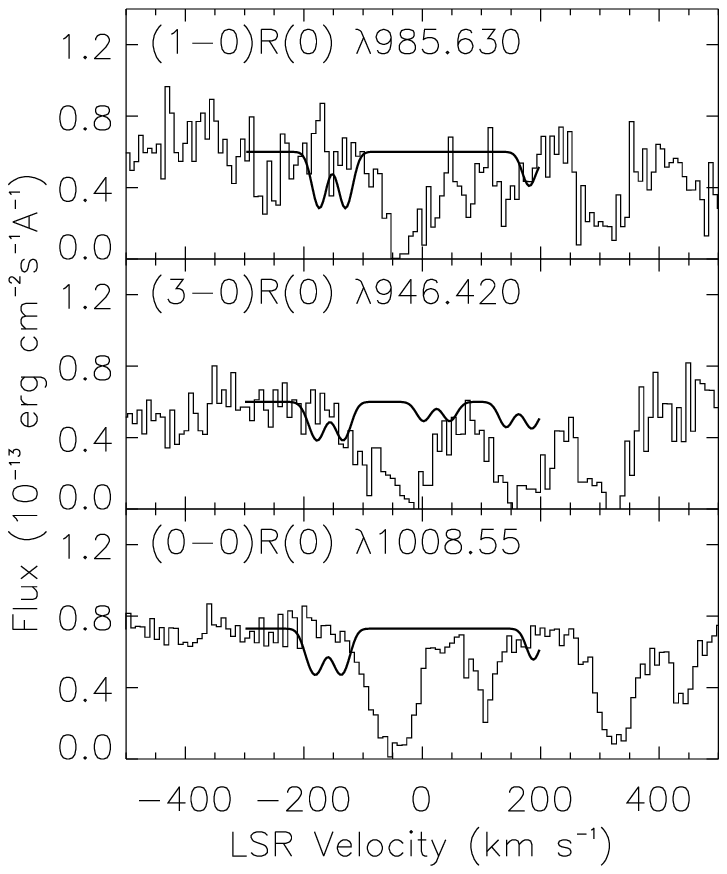}
\caption{A comparison of the observed spectrum with
expected absorption signatures of two components of \mh\ with 
$\vlsr=-172\;\kms$ and $\vlsr=-133\;\kms$ each having
N(\mh)$=2.0\times10^{14}\;\cmm{-2}$ for a temperature T$_{01}=100$ K and
$b=10\;\kms$.  If \mh\ were present at this column
density, it would have been observed.
\label{fig2}}

\end{figure*}

Previous searches for molecular gas in high velocity clouds
concentrated on tracers such as CO and HCO$^{+}$, which have spectral
lines at millimeter wavelengths.  \citet{wakker97} and
\citet{akeson99} placed upper limits on N(\mh) of $9\times10^{18}
\cmm{-2}$ and $5.5\times10^{18} \cmm{-2}$ in several HVCs,
respectively, assuming solar abundances and standard I(CO) to N(\mh)
ratios.  \citet{richter99} have reported the discovery of 
\mh\ absorption at a velocity of +120 \kms\ with
N(\mh)=$(2.2-3.6)\times10^{15}\;\cmm{-2}$ 
in an {\it ORFEUS} spectrum of the LMC star HD 269546
(Sk$-$68\,82).  However, this sightline is  quite complicated, and
has absorption associated with the Milky Way, the LMC, as well as the
HVC.

{\it FUSE} can observe the far UV electronic transitions of molecular
hydrogen from both the Lyman (B--X) and Werner (C--X) bands at high
resolution toward many extragalactic objects behind HVCs.
Toward Mrk 876, absorption lines from \mh\ in the
Milky Way are readily visible with
N(\mh)$\sim2.3\times10^{18}\;\cmm{-2}$ \citep{shull00}.  
Figure 2 compares  
the observed spectra and the expected absorption signatures of the two
components of complex~C for an \mh\ column density of
$2\times10^{14}\;\cmm{-2}$ in each component and for a rotational
temperature of T$_{01}=100$ K and $b=10\;\kms$.  The selected lines are the
strongest \mh\ lines in the {\it FUSE} bandpass that are not blended with
other atomic or molecular lines.  Clearly, there is no detectable \mh\
absorption from HVC complex~C.  Reasonable changes in T$_{01}$ or $b$
do not alter this conclusion.  Since the most efficient mechanism for
forming \mh\ is on the surfaces of dust grains \citep{spitzer78}, 
the non-detection of
\mh, coupled with the low depletion of iron, implies that there
is little or no dust in complex~C along this line of sight.

\begin{table*}
\epsscale{2.0}
\plotone{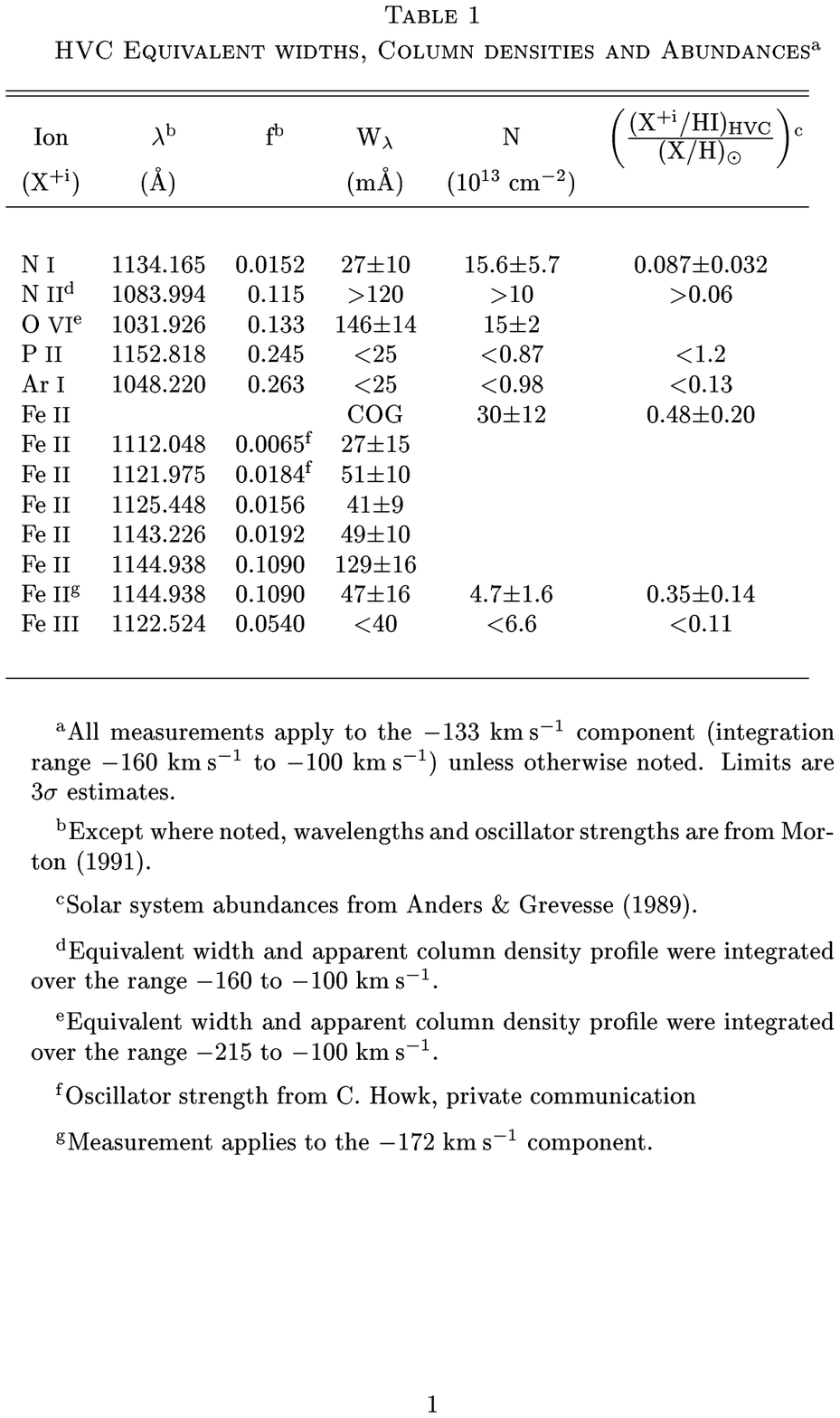}
\end{table*}

\acknowledgments

This work is based on data obtained for the Guaranteed Time Team by
the NASA-CNES-CSA {\it FUSE} mission operated by the Johns Hopkins
University.  Financial support to U.S. participants has been provided
by NASA contract NAS5-32985.  We thank Peter Kalberla for
observing the Effelsberg \HI\ spectrum. The Effelsberg Telescope is
operated by the Max Planck Institut f\"ur Radio Astronomie at Bonn.
We also thank Matt Haffner for providing the WHAM H$\alpha$ result.

%
%
%
%


\begin{thebibliography}{}
\bibitem[Akeson \& Blitz(1999)]{akeson99}%
         Akeson, R.L., Blitz, L. 1999, \apj, 523, 163
\bibitem[Anders \& Grevesse (1989)]{anders89}%
         Anders, N., Grevesse, E. 1989, Geochem. Cosmochim. Acta, 53, 1973
\bibitem[Gibson et al.\ (2000)]{gibson00}%
         Gibson, B.K., Giroux, M.L., Penton, S.V., Stocke, J.T., Shull, J.M. 
         2000, \aj, submitted
\bibitem[Moos et al.\ (2000)]{moos00}%
         Moos, H.W. et al. 2000, \apjl, this volume.
\bibitem[Morton(1991)]{morton91}%
         Morton, D.C. 1991 \apjs, 77, 119
\bibitem[Reynolds et al.\ (1998)]{reynolds98}%
         Reynolds, R.J., Tufte, S.L., Haffner, L.M., Jaehnig, K., Percival, J.W.
 1998, Proc.\ Astr.\ Soc.\ Austr., 15, 14
\bibitem[Richter et al.\ (1999)]{richter99}%
         Richter, K.S., de Boer, K.S., Widmann, H., Kappelmann, N., Gringel, W.,
 Grewing, M., Barnstedt, J. 1999, \nat, 402, 386 
\bibitem[Sahnow et al.(2000)]{sahnow00} Sahnow, D. J., et al. 2000, \apjl, this volume.
\bibitem[Savage \& Sembach (1996)]{savage96}%
         Savage, B.D., Sembach, K.S. 1996, \araa, 34, 279
\bibitem[Sembach(1999)]{sembach99}%
         Sembach, K. S. 1999, PASP Conf. Ser. 166, 243
\bibitem[Sembach et al.\ (2000)]{sembach00}%
         Sembach, K.S., et al. 2000, \apjl, this volume
\bibitem[Shull et al.(2000)]{shull00} Shull, J. M., Tumlinson, J.,
        et al. 2000, \apjl, this volume
\bibitem[Sofia \& Jenkins (1998)]{sofia98}%
         Sofia, U.J., Jenkins, E.B. 1998, \apj, 499, 951
\bibitem[Spitzer(1978)]{spitzer78}%
         Spitzer, L. 1978, Physical Processes in the Interstellar
Medium (John Wiley \& Sons, New York)
\bibitem[Sutherland \& Dopita (1993)]{sutherland93}%
         Sutherland, R.S., Dopita, M.A. 1993, \apjs, 88, 253
\bibitem[van Woerden et al.\ (1999)]{vanwoerden99}%
         van Woerden, H., Peletier, R. F., Schwarz, U. J., Wakker, B. P., 
         \& Kalberla, P. M. 1999, PASP Conf. Ser. 166, 1
\bibitem[Wakker et al.\ (1997)]{wakker97}%
         Wakker, B.P., Murphy, E.M.,  van Woerden, H. Dame, T. 1997, \apj, 
         488, 216 
\bibitem[Wakker \& van Woerden (1991)]{wakwoer91}%
         Wakker, B.P. \& van Woerden, H. 1991, \aap, 250, 509
\bibitem[Wakker \& van Woerden (1997)]{wakwoer97}%
         Wakker, B.P. \& van Woerden, H. 1997, \araa, 35, 217
\bibitem[Wakker et al.\ (1999)]{wakker99}%
         Wakker, B.P., Howk, J.C., Savage, B.D., van Woerden, H., Tufte, S.L., 
         Schwarz, U.J., Benjamin, R., Reynolds, R.J., Peletier, R.F., 
         Kalberla, P.M.W. 1999, \nat, 402, 388

\end{thebibliography}
\end{document}